\documentclass[a4paper,12pt]{article}
\usepackage{latexsym,amssymb}
\usepackage{amsmath,amssymb,amsxtra,amscd,amsthm}
\usepackage{amsfonts}
\usepackage[mathscr]{eucal}
\DeclareFontFamily{U}{rsfs}{\skewchar\font127 }
\DeclareFontShape{U}{rsfs}{m}{n}{
   <5> rsfs5
   <6> rsfs6
   <7> rsfs7
   <8> rsfs8
   <9> rsfs9
   <10> rsfs10
   <10.95> rsfs11
   <12> rsfs12
   <14.4> rsfs14
   <17.28> rsfs17
   <20.74> rsfs20
   <24.88> rsfs25
   <29.86-> rsfs30}{}
\DeclareMathAlphabet\scr{U}{rsfs}{m}{n}

\def\mZ{\mathbb{Z}}

\def\tL{\mathrm{L}}

\def\tS{\mathrm{S}}

\DeclareFontFamily{U}{rsf}{} \DeclareFontShape{U}{rsf}{m}{n}{
  <5> <6> rsfs5 <7> <8> <9> rsfs7 <10-> rsfs10}{}
\DeclareMathAlphabet\Scr{U}{rsf}{m}{n}

\def\bb1{\textup{\small{1}} \kern-3.8pt \textup{1}}

\def\SL2Z{\tS\tL(2,\mZ)}

\numberwithin{equation}{section}

\providecommand{\href}[2]{#2}

\newtheoremstyle{plain} 
  {5pt}
  {10pt}
  {\rmfamily} 
  {}
  {\scshape} 
  {}
  {\newline} 
  {}
\swapnumbers

\def\a{\alpha}
\def\b{\beta}
\def\g{\gamma}
\def\l{\lambda}

\begin{document}

\begin{titlepage}
\begin{flushright}
DISTA-2010 \\
hep-th/yymmnnn
\end{flushright}
\vskip 1.5cm
\begin{center}

{\LARGE \bf Pure Spinor Integration \\ \vspace{.2cm }from the Collating Formula}
 \vfill
\vskip 0.5cm
{\large  P. A. Grassi and L. Sommovigo}
\vfill 
{
{ DISTA, Universit\`a del Piemonte Orientale, }\\
{ Via Teresa Michel 11,  Alessandria, 15121, Italy
$\&$ INFN - Sezione di Torino
\vskip .1cm
{\small \sf pietro.grassi@mfn.unipmn.it\,, ~~~~ luca.sommovigo@mfn.unipmn.it
}
}}
\end{center}
\vfill
\begin{abstract}

We use the technique developed by  Becchi and Imbimbo to construct a
well-defined BRST-invariant path integral formulation of pure spinor
amplitudes. The space of pure spinors can be viewed from the algebraic
geometry point of view as a collection of open sets where the
constraints can be solved and a free independent set of variables can
be defined. On the intersections of those open sets, the functional
measure jumps and one has to add boundary terms to construct a
well-defined path integral. The result is the definition of the pure
spinor integration measure constructed in term of differential forms
on each single patch.
\end{abstract}
\vfill

\end{titlepage}

\section{Introduction}

One of the main ingredients in the computation of superstring
amplitudes is the construction of a path integral measure for the pure
spinors \cite{Berkovits:2004px, Berkovits:2006vi,Bedoya:2009np} (in
the text we denote by $\lambda^\a$ the 10d pure spinors satisfying the
algebraic equation $\l \g^m \l =0$. They are chiral worldsheet bosons
and they appear holomorphically in the action and in the path
integral). 

In the functional integration, we have to distinguish
between the integration over the zero modes and the integration over
the non-zero modes. The latter can be performed by using the OPE
technique and computing the contractions between vertex operators. The 
former, however, requires much more care because of the pure spinor
condition. Actually, the pure spinor space can be seen as an
algebraic subvariety of the complex space ${\mathbb C}^{16}$ and,
therefore, some techniques of algebraic geometry may be used to
construct the amplitudes (see for example
\cite{Nekrasov:2005wg,Berkovits:2006vi,Gomez:2009qd,Gomez:2010ad}). A
fundamental observation is the existence of a nowhere-vanishing
holomorphic top-form which has been constructed in term of the pure
spinor variables \cite{Nekrasov:2005wg}. In terms of such form one can
define the integration on the pure spinor space and the physical
amplitudes.

However, this integration is subject to some drawbacks: first, the
integration on the pure spinor variables is a holomorphic integration
(see for example the discussion in
\cite{Hoogeveen:2009hk,Grassi:2009fe}) and therefore a holomorphic
curve should be given in the pure spinor space (and so far nobody has
provided a construction for that path). Nevertheless, in several
cases, the algebraic properties of the pure spinors -- they carry a
representation of the Poincar\'e group -- are sufficient to determine
the value of the integration without defining such a
curve.\footnote{See also \cite{Lazaroiu:2003vh}. An analogous problem
  appears in topological strings and holomorphic matrix integrals.}
Second, the space is not compact and it requires a regularization in
order to perform meaningful operations.   

One way to solve these problems is to introduce new coordinates (in
the text we will denote them by $\bar \l_\a$), leading to the
non-minimal formalism \cite{Berkovits:2005bt} where the pure spinor
integration is seen as a real integration, as recently discussed in
\cite{Gomez:2009qd,Gomez:2010ad}. Indeed, this easily yields the
correct coefficients for one- and two-loop amplitudes.

On the contrary, by considering the new variables $\bar\lambda_\a$ as
conjugated to the pure spinor variables (satisfying themselves the
pure spinor conditions), one can adopt a different point of view: the idea is to
construct the pure spinor integration measure from cohomology theory
using the isomorphism between the \v Cech cohomology of the pure
spinor space and the de Rham cohomology. For that purpose one needs a
partition of unity -- which depends upon the conjugated variables--, 
the measure on a single patch and the {\it
  Collating Formula} \cite{bott-tu}, which provides the correct
machinery to build globally defined forms on a space described in term
of its pacthes. Starting from a $p$-form on a given patch and written
in terms of local coordinates, one can obtain a corresponding globally
defined $p$-form on the entire space. The result is not holomorphic,
since the partition of unity is not holomorphic, nevertheless it
yields a global top-form to be used to compute the amplitudes. In
addition, it may provide a different way to regularize  the behavior
of the measure at infinity by introducing suitable cut-offs on the
patches. This  might be very useful to avoid some problems pointed out
in literature
\cite{Berkovits:2006vi,Grassi:2009fe,Aisaka:2009yp,Park:2010xw,Bjornsson:2010wm}.  

The collating formula is a purely mathematical result and its
derivation is independent of string theory or quantum field
theory. Nonetheless, Becchi and Imbimbo \cite{Becchi:1995ik} derived
it from a path integral approach analyzing the boundary terms needed
to define global forms on punctured Riemann surfaces. The path
integral approach shows how the jumps in the measure passing from one
patch to another yield boundary terms which can be removed by adding
contributions to the measure reproducing the collating formula. 

In the present paper, we do not discuss the applications of the present 
formalism and they will be postponed in a subsequent publication dedicated 
to the amplitude computations with the present formula.

In sec. 2 we present a construction of the measure for algebraic
varieties by generalizing the Griffiths' method in the case of
non-complete intersections. In addition, we derive the measure in a
different way resulting in a vector-valued holomorphic form and we
explain the relation with the holomorphic top form. At the end of the
section, we construct a real globally defined form. In sec. 3 we
discuss the mathematical Collating Formula. In sec. 4 we translate it
in term of a BRST formulation and a path integral technique leading to
a globally defined real form. In sec. 5 we provide some examples of the application 
of the collating formula. 

\section{Integration on Constrained Spaces}

We consider a set of variables $Z_i$ constrained by the algebraic
equations $W^a(Z)=0$. They describe  the $k$-dimensional hypersurfaces
denoted by ${\cal P}$. The index $a=1,\dots,q$ runs over the number of
polynomials $W^a(Z)$ in the variables $Z_i$ and $i$ runs over the
dimension of the ambient manifold which is assumed to be ${\mathbb
  C}^{N}$.

If the space is a complete intersection, the constraints $W^a(Z)$ are
linearly independent and the differential form
\begin{equation}\label{ICSA}
\Theta^{(N-k)}= \epsilon_{a_1 \dots a_{N-k}} dW^{a_1}\wedge \dots
\wedge dW^{a_{N-k}}\,, 
\end{equation}
is not vanishing. In this case, $q=N-k$ and the dimension of the
surface is easily determined. For example, if the hypersurface is
described by a single algebraic equation $W(Z)=0$, the form
(\ref{ICSA}) is given by $\Theta^{(1)} = d \, W$. 

On the other hand, if the hypersurface is not a complete intersection,
then there exists a differential form 
\begin{equation}\label{ICSB}
  \Theta^{(N-k)}= {\cal T}_{A, [a_1 \dots a_{q}] } dW^{a_1}\wedge
  \dots \wedge dW^{a_{q}} \wedge \eta^{A,(N-k-q)}\,,  
\end{equation} 
where $ \eta^{A,(N-k-q)}$ is a set of $N-k-q$ forms defined such that
$\Theta^{(N-k)}$ is non-vanishing on the constraints $W^a(z)=0$ and
${\cal T}_{A, [a_1 \dots a_{q}] }$ is a numerical tensor which is
antisymmetric in the indices $a_1\dots a_q$. The construction of $
\eta^{A,(N-k-q)}$ depends upon the precise form of the algebraic
variety. In some cases a general form can be given, but in general it
is not easy to find it and we did not find a general procedure 
for that computation.\footnote{In the case of Normal Rational Curves
  (which are not complete intersections), the form is given by
  $\Theta^{(N-2)} = \epsilon^{I_1 \dots I_q I_{q+1} \dots I_N}
  dW_{I_1} \wedge \dots \wedge dW_{I_q} \wedge \omega^1_{I_{q+1}}
  \wedge \dots \wedge \omega^{N-q}_{I_{N}}$ where the vectors
  $\omega^{r}$ are defined by means of the relations (syzygy) between
  the constraints $W^I$ such that $\omega^r_I(Z) W^I(Z) =0$ for all
  $r=1,\dots,N-q$. In this case the vectors $\omega^r_I(Z)$ are global
  functions on the space.}  

To construct a global form on the space ${\cal P}$ one can use a
modification of the Griffiths' residue method \cite{griffiths-harris} 
by observing that given the global holomorphic form on the ambient
space $\Omega^{(N)} = \epsilon^{i_1 \dots i_N} dZ_{i_1} \wedge \dots
\wedge dZ_{i_N}$, we can decompose the $\{ Z_i\}$'s into a set of
coordinates $Y^a= W^a(Z)$ and the rest. 
By using the contraction with respect to $q$ vectors $\{\bar
Z^a_{i}\}$, the top form for ${\cal P}$ can be written as   
\begin{equation}\label{ICSC}
  \Omega^{(k)} = \frac{\iota_{\bar Z^{a_1}} \dots \iota_{\bar Z^{a_q}} 
  \Omega^{(N)}}{\iota_{\bar Z^{a_1}} \dots \iota_{\bar Z^{a_q}}
  \Theta^{(N-k)}}  
\end{equation}
which is independent from $\{\bar Z^a_{i}\}$  as can be easily proved
by using the constraints $W^a(Z)=0$. Notice that this form is
nowhere-vanishing  and non singular only in the case of CY space. The
vectors $\{\bar Z^a_{i}\}$ play the role of gauge fixing parameters
needed to choose a polarization of the space ${\cal P}$ into the
ambient space.  

In the case of pure spinor we have: the ambient form $\Omega^{(16)} =
\epsilon_{\a_1\dots \a_{16}} \, d\lambda^{\alpha_1} \wedge \dots \dots
\wedge d\lambda^{\alpha_16}$ and $\Theta^{(5)} = \lambda \g^m d\lambda
\, \lambda \g^m d\lambda\, \lambda \g^m d\lambda\, d\lambda \g_{mnp}
d\lambda$. From these data, we can get the holomorphic top form
$\Omega^{(11)}$ by introducing 5 independent parameters $\bar\lambda$
and by using the formula (\ref{ICSC}). The latter is independent from
the choice of the parameters $\bar\lambda$ (however, some care has to
be devoted to the choice of the contour of integration and of the
integrand: in the minimal formalism, the presence of delta functions
$\delta(\lambda)$ might introduce some singularities which prevent
from proving the independence from $\bar\lambda$, as was pointed out
in \cite{Hoogeveen:2009hk,Berkovits:2009hp}).

Using $\Omega^{(k)} \wedge \overline{\Omega^{(k)}}$, 
one can compute the correlation functions by
integrating globally defined functions. When the space is Calabi-Yau,
it also exists a globally-defined nowhere-vanishing holomorphic form
$\Omega^{(k|0)}_{hol}$ such that $\Omega^{(k|0)}_{hol} \wedge
\overline{\Omega^{(0|k)}_{hol}}$ is proportional to
$\Omega^{(k)} \wedge \overline{\Omega^{(k)}}$. The ratio of the two top forms is a globally defined
function on the CY space.  

In the case of the holomorphic measure $\Omega^{(k|0)}_{hol}$ 
the integration of holomorphic functions is related to the
definition of a contour $\gamma \in {\cal P}$ in the complex space  
\begin{equation}\label{intA}
\langle \prod_A {\cal O}(p_A) \rangle = \int_{\gamma \in \cal P}
\Omega^{(k,0)}(Z_i) \prod_A {\cal O}_0(Z_i, p_A)
\end{equation}
where ${\cal O}(Z_i, p_A)$ are the vertex operators of the theory
localized at the points $p_A$ of the Riemann surface and ${\cal
  O}_0(Z_i, p_A)$ is the zero-mode component of the vertex
operators. They are functions of the coordinates $Z_i$ which are
reduced to their zero modes (the non-zero mode part is computed by the
usual OPE technique). 

For instance, let us consider the hypersurface $\sum_i Z_i^2 =0$ in
${\mathbb C}^N$. This equation can be put in the form $P(u_{i'}) = w
z$ where $i'$ runs over $i'=1,\dots,N-2$ coordinates and $w,z$ are two
combinations of the $Z$'s. $P(u_{i'})$ is a polynomial of the
coordinates $u_{i'}$. For a given N they are local CY spaces and
there exists a globally-defined, nowhere-vanishing holomorphic top
form given by 
\begin{equation}\label{intB}
\Omega^{(N-1|0)} = \frac{dz \wedge du_{i^1}\wedge\dots \wedge
  du_{i^{N-2}}}{z} 
\end{equation}
In this case, the ambient form is $\Omega^{(N)} = \epsilon_{i_1 \dots
  i_N} d Z^{i_1}\wedge\dots \wedge d Z^{i_{N}}$ while the $\Theta$
form is $\Theta^{(1)} = \sum_{i=1}^N Z^i d Z^i$. Again, by introducing
the parameters $\bar Z$ (and assuming that they transform under a
vector representation of $SO(N)$), this form can also be written in an 
$SO(N)$ invariant way by using (\ref{ICSC}) 
\begin{equation}\label{intC}
\Omega^{(N-1,0)} = \frac{\bar Z_{i_1} \epsilon^{i_1 \dots i_N}
  dZ_{i_2}\wedge\dots \wedge dZ_{i_{N}}}{ \bar Z^i Z_i} 
\end{equation}
Note from (\ref{intB}) that the measure has a possible pole for $z=0$,
however the form of (\ref{intB}) depends upon the choice of $\bar Z$
and this corresponds to the choice of a patch where the corresponding
coordinate does not vanish. Since, in general (as discussed after
(\ref{ICSC})) the measure is independent from the $\bar Z$, also the
measure given in (\ref{intB}) is free from singularities. 

There is another way to construct a measure for these examples which
is independent of the choice of the $\bar Z$, namely a measure with
values in the tangent vector bundle.  
We define the tensor $T^{(1,N-1|0,0)}$ which is a holomorphic vector
with values in the $N-1$-form space as follows 
\begin{equation}\label{intD}
T^{(1,N-1|0,0)} = \epsilon^{i_1 \dots i_N} dZ_{i_1} \wedge \dots
\wedge dZ_{i_{N-1}} (g^{-1})_{i_N k}\partial^k 
\end{equation}
which is compatible with a single constraint $W(Z_i)=0$ and 
\begin{equation}\label{intDA}
 g^{l k} = \frac{\partial^2 W}{\partial Z_{l} \partial Z_k} \,. 
\end{equation}
The relation between (\ref{intC}) and (\ref{intD}) is given by the
equation 
\begin{equation}\label{intDB}
\Omega^{(N-1|0)} = T^{(1,N-1|0,0)}\left[ \ln\left( \bar Z^i g_{ij} Z^j
  \right)\right]\,,
\end{equation}
where the differential operator acts on the globally defined
function $ \ln\left( \bar Z^i g_{ij} Z^j \right)$, which is singular
only for $\bar Z^i g_{ij} Z^j  =0$. In the present case there is a
single constraint and therefore a single set of parameters $\bar Z^i$
and we identify them with the complex conjugated to $Z^i$.

In the case of pure spinors the measure is given by 
\begin{equation}\label{intDC}
T^{(3,11|0,0)} = \epsilon_{\a_1\dots \a_{16}} d\lambda^{\a_1} \wedge
\dots \wedge d\lambda^{\a_{11}}(\g^{mnp})^{\a_{12} \a_13} (\g_m
\partial_{\lambda})^{\a_{14}} (\g_n \partial_{\lambda})^{\a_{16}}
(\g_p \partial_{\lambda})^{\a_{16}}\,. 
\end{equation}
which is covariant under Lorentz transformations and, acting on the
function $\ln (\bar \lambda_\alpha \lambda^\a)$, it is related to the
expression obtained in the non-minimal formalism \cite{Berkovits:2005bt}.  

Until now, we have constructed global holomorphic measures for the
integration on the zero modes. However, as we have already pointed
out, the integration over such a measure requires a holomorphic
curve. Nevertheless we can avoid such a trouble by constructing a real
measure (which amounts to a specific choice of integration countour).
We observe that, if we define the measure on the conjugated variables
using the rule (\ref{intB}) 
\begin{equation}\label{intE}
\overline \Omega^{(0,0|0,N-1)} = \frac{Z_{i_1} \epsilon^{i_1 \dots
    i_N} d\bar Z_{i_2} \wedge \dots \wedge d\bar Z_{i_{N}}}{\bar Z
  \cdot Z} 
\end{equation}
where we used the variables $Z^i$ as gauge parameters in (\ref{intB}),  
we can act on it with the differential operator $T^{(1,N-1|0,0)}$ to
get  
\begin{equation}\label{intF}
T^{(1,N-1|0,0)}\wedge\left[ \overline \Omega^{(0,0|0,N-1)}\right] =
dZ_{i_2} \wedge \dots \wedge dZ_{i_{N}} \epsilon^{i_1 \dots i_N}
G_{i_1 j_1} \epsilon^{j_1 \dots j_N} d\bar Z_{j_2} \wedge \dots \wedge
d\bar Z_{i_{N}}  
\end{equation}
where 
\begin{equation}\label{intG}
G_{i_1 j_1} = \left( \frac{\delta_{i_1 j_1} \bar Z\cdot Z  - \bar
    Z_{i_1} Z_{j_1}}{(\bar Z \cdot Z)^2} \right) 
\end{equation}
is the Fubini-Study metric on ${\mathbb P}^{N-1}$ written in terms of
the homogenous coordinates.  
A similar result has been obtained by Gomez in \cite{Gomez:2009qd}. 
This measure has the advantage of being real and globally defined, but
it is clearly not holomorphic and it does not require a path to be
specified.

This construction of the measures does not make use of the
decomposition of the manifold in patches but, if we are willing to
abandon the holomorphicity, we can build the top forms from a
completely different point of view. 
Namely, we start from the measure on a single patch, which is the
usual flat measure written in terms of a set of convenient
coordinates, and we construct the global measure using the {\it
  Collating Formula} by gluing the contributions coming from the
different patches.  

\section{Collating formula}
A fundamental theorem in cohomological theory is \cite{bott-tu}   

{\it Theorem}: If ${\mathfrak U}$ is a good cover of the manifold
${\cal M}$, then the de Rham cohomology $H^*_{DR}({\cal M}) $ of
${\cal M}$ is isomorphic to the \v Cech  cohomology $H^*({\mathfrak
  U}, {\mathbb R})$ of the good cover  
\begin{equation}\label{collA}
H^*_{DR}({\cal M}) \simeq H^*({\mathfrak U}, {\mathbb R})\,,
\end{equation}
where the inclusions 
\begin{equation}\label{collB}
{\cal M} \leftarrow {\cal U}_I \,
\raisebox{-2pt}{$\leftarrow$}
\hspace{-0.42cm} \raisebox{2pt}{$\leftarrow$} \, \,
{\cal U}_{IJ} \, 
\raisebox{-3pt}{$\leftarrow$}
\hspace{-0.54cm} \leftarrow
\hspace{-0.54cm} \raisebox{3pt}{$\leftarrow$}
\,\,\,
{\cal U}_{IJK} \, \dots
\end{equation}
include ${\cal U}_I$,  the open sets of the good cover into the
multiple intersections of the open sets ${\cal U}_{I_1 \dots I_n}
\equiv \underset{i=1 \dots n}{\cap} {\cal{U}}_{I_i}$. 

The meaning of the present theorem is the following: on one hand,
the differential geometry of forms establishes an exact sequence in
the complex $0 \rightarrow \Omega^*({\cal M})
\stackrel{r}{\rightarrow} C^*({\mathfrak U}, \Omega^*)$, where $r$ is
the restriction map which restricts the exterior algebra
$\Omega^*({\cal M})$ to the complex of cochains of differential forms
$C^*({\mathfrak U}, \Omega^*) = \oplus_{p,q\geq 0} C^p({\mathfrak U},
\Omega^q)$ on the cover of ${\cal M}$. 
On the other hand, starting from the combinatorics of the cover
${\mathfrak U}$, one can compute the Mayer-Vietoris sequence of 
the complex $0 \rightarrow C^*({\mathfrak U}, {\mathbb R}) \rightarrow
C^*({\mathfrak U}, \Omega^*)$. Finally, in the double complex the two
cohomologies are mixed. In this way, one proves that $H^*_{DR}({\cal
  M}) \simeq H_D(C^*({\mathfrak U}, \Omega^*))$ using the fact that
the de Rham cohomology of ${\cal M}$ is isomorphic to the cohomology
of the double complex. In addition, if ${\mathfrak U}$ is a good
cover, the \v Cech cohomology is also isomorphic to the cohomology of
the double complex $H^*({\cal M},{\mathbb R}) \simeq
H_D(C^*({\mathfrak U}, \Omega^*))$, and therefore it follows the
isomorphism between the de Rham and the \v Cech cohomology.  

The main point is that this isomorphism provides a way to compute the
de Rham cohomology by means of combinatorics of a good cover (we refer
to \cite{bott-tu} for a complete discussion regarding the existence of
good cover and its implications). Here we only describe the crucial
formula providing the explicit isomorphisms between the two
cohomologies.  
Given $f$ a chain map
\begin{equation}\label{collC}
f: C^*({\mathfrak U}, \Omega^*) \rightarrow \Omega^*({\cal M})
\end{equation}
such that $f \circ r =1$ and $r \circ f$ is chain homotopic to the
identity. The map $f$ provides the main ingredient for collating
together a \v Cech-de Rham cochain into a global form. This is given
by the {\it Collating Formula}. For the description of the collating
formula we need to define a homotopy operator $K$ by introducing the
partition of unity $\rho_I$ (such that every point of ${\cal M}$ has a
neighborhood in which $\sum_I \rho_I$ is a finite sum, $\sum_I \rho_I
=1$ and defined such that ${\rm supp}[{\cal U}_I] \subset {\cal U}_I)$
and $K$ acting on a $p$-cochain $\omega_{I_0 \dots I_{p}}$ gives 
\begin{equation}\label{collD}
(K \omega)_{I_0 \dots I_{p-1}} = \sum_I \rho_I \omega_{I I_0 \dots
  I_{p-1}}\,. 
\end{equation}
If $\delta$ is the \v Cech operator $(\delta \omega)_{I_0 \dots I_{p}}
= \sum_{i=1}^{p} (-)^i \omega_{I_0\dots \hat I_i \dots I_p}$ (where
hatted index must be omitted), then the homotopy operator $K$
satisfies the relation $K \delta + \delta K = 1$. 
The last ingredients are the differential operators $D''= (-1)^p
d$, acting on the complex $C^*({\mathfrak U}, \Omega^*)$ where $p$ is
the degree of the cochain, and $D = \delta + D''$.

With these ingredients we can provide the Collating Formula. 
Let $K$ be the homotopy operator defined above and $\a = \sum_{i=0}^n
\a_i$ is an $n$-cochain such that $D \a = \beta = \sum_{i=0}^{n+1}
\beta_i$.  
They satisfy the following descent equations 
\begin{eqnarray}\label{collDA}
d \alpha^{p}_{0} = \beta^{p+1}_{0}\,, \quad
d \alpha^{p-1}_{1} + \delta \alpha^{p}_{0} = \beta^{p}_{1}\,, \quad\dots\quad
d \alpha^{0}_{p} + \delta \alpha^{1}_{p-1} = \beta^{1}_{p}\,, \quad
\delta \alpha^{0}_{p} = \beta^{0}_{p+1}\,, \quad
\end{eqnarray}
where the differential operator $d$ increases the form number and the
difference operator $\delta$ increases the co-chain number. If we act
with the homotopy operator $K$ on the last element of the descent
equation, we get $K \, \delta  \alpha^{0}_{p} = K\, \beta^{0}_{p+1}$,
and using the algebraic relation between K and $\delta$, it yields
$\alpha^{0}_{p} = \delta K \alpha^{0}_{p} + K\, \beta^{0}_{p+1}$. In
the same way, inserting this result in the next-to-the-last equation
in (\ref{collDA}), we obtain the next cochain $\alpha^{1}_{p-1}$ in
terms of $\beta$'s and $\alpha^{0}_p$. Proceeding in this way we get
the following final formula 
\begin{equation}\label{collE}
f(\a) = \sum_{i=0}^{n} (- D'' K)^i \alpha_i - \sum_{i=1}^{n+1} K ( - D'' K)^{i-1} \beta_i \,\, \in 
C^0({\mathfrak U}, \Omega^n)\,.
\end{equation}
This is a global form satisfying the above mentioned properties. We
refer to \cite{bott-tu} for the proofs and the discussion. In the
present form, this formula is not very useful for our purposes since
we do not know the various terms and they must be computed from
(\ref{collE}). For that purpose it is convenient a path-integral
derivation adapted to the problems discussed in the previous
section. In that way the all ingredients in the above formula are
specified and the global form can be easily derived.\footnote{In
  \cite{Becchi:1995ik} the solution of the descent equation is
  obtained by introducing a set of anticommuting auxiliary variables
  $\xi^\alpha$ and rewriting the \v Cech operator as $\delta =
  \sum_{\alpha} \xi^{\alpha}$ and the homotopy operator as $K =
  \sum_{\alpha} \rho_{\alpha} \partial_{\xi^\alpha}$.}

The main idea is that the path integral has a jump passing from one
patch to another and this jump is seen as an anomaly in the Ward
Identity. The contributions needed to remove such anomaly terms are
indeed the addends in (\ref{collE}).

\section{(Path Integral) Collating Formula}

We translate the above algebraic-differential derivation into a path
integral formulation adapted to sigma models and string models.  

Our model is defined by the action $S[Z] = \int_{\Sigma} d^2z {\cal
  L}(Z(z,\bar z))$ where ${\cal L}(Z(z,\bar z))$ is the Lagrangian of
the sigma model and  $Z(z,\bar z): \Sigma \rightarrow {\cal M}$ maps
the worldsheet $\Sigma$ into the target space ${\cal M}$.  
We assume that the target space is a complex manifold with a K\"ahler
form $K = K_{i \bar j} \, dZ^i \wedge d\bar Z^{\bar j}$, where $\bar
Z^{\bar i}$ are the complex conjugated variables. 
The action needs not be real and we assume to be starting with a
holomorphic Lagrangian ${\cal L}(Z)$; the gauge-fixing will possibly
involve the complex conjugated coordinates. 

In general, the target space ${\cal M}$ can be better described in
terms of an atlas, namely in terms of a given set of open sets ${\cal
  U}_I$ and a system of coordinates $Z_{(I)}$'s on each of them. The
form of the target space is parameterized by the transition functions
between the different patches 
\begin{equation}\label{transA}
h_{JI} : ({\mathcal U}_I, Z_{(I)}) \longrightarrow ({\mathcal U}_J,
Z_{(J)})  
\end{equation}
such that $h_{JI}( {\mathcal U}_I) \subset {\mathcal U}_J$ and
$h_{JI}(Z_{(I)}) = Z_{(J)}$.
We assume that the transition functions $h_{IJ}$ are holomorphic
functions of $Z$'s. (A partition of unity for the atlas $\bigcup_I
(\mathcal{U}_I , Z_{(I)})$ can be given by $\rho_{I} = \frac{\bar Z_I
  Z_I}{ \bar Z \cdot Z}$, with $\bar Z \cdot Z = \bar Z^i Z_i$, and
where we have assumed that on the patch $I$ the coodinate $Z_i$ with
$i = I$ is different from zero).

The local observables of the theory are given by the vertex operators
${\mathcal O}_A(Z)$ (where $A$ runs over the whole set of vertex
operators) and they must be globally defined on ${\cal M}$,
holomorphic and BRST invariant.  

The correlation functions are computed as follows
\begin{align}\label{transB}
&\Big\langle \prod_A {\cal O}_A(Z) \Big \rangle = \prod_A
\frac{\delta}{\delta \eta_A}  
\left. Z[\eta_A] \right|_{\eta=0}\nonumber \\
& Z[\eta_A] =  \int {\cal D}Z e^{-S[Z] + \sum_A \eta_A {\cal O}_A(Z)} 
\end{align}
$\eta_A$ are the sources for the observables and ${\cal{D}}Z$ is the
naive functional integration measure. 
This formula is intrinsically badly-defined if there is not a global
coordinate system, since the naive integration measure ${\cal{D}}Z$
makes sense only in the open patches $\cal{U}_I$. 
Moving from one patch to another, the integration measure could
receive contributions from the intersections of the different patches
and these contributions are fundamental to provide a functional
measure which is globally defined on the whole target space.

Therefore, we start from a different expression 
\begin{align}\label{transC}
Z[\eta_A] =  \sum_I \int {\cal D}Z_{(I)} e^{-S[Z_{(I)}] + \sum_A
  \eta_A {\cal O}_A(Z_{(I)})} 
\end{align}
where the integration is done patch by patch and we use the naive
integration on any single patch ${\cal D}Z_{(I)}$. If we decompose the
field $Z_{(I)}$ in terms of zero modes and non-zero modes, the naive
functional integration measure is given by  
\begin{equation}\label{transCA}
{\cal D}Z_{(I)} = \Big|dZ^1_{(I) 0} \wedge \dots \wedge  dZ^n_{(I) 0}
\prod_{k=1}^{\infty} dZ^1_{(I) k} \wedge \dots \wedge  dZ^n_{(I)
  k}\Big|^2.  
\end{equation}
The observables ${\cal O}_A(Z_{(I)})$ are defined by means of the
restriction map (see sec. 3) from the globally defined observables
${\cal O}_A(Z)$ and the action $S[Z_{(I)}]$ is defined by the
restriction map from the action $S[Z]$.

Now, we assume that the non-zero modes are treated with the
conventional OPE technique, so that we can consider only the zero
modes of the theory.
We write the integrand as an $(n|n)$ form
\begin{equation}\label{rapA}
\omega^{(n|n)}_I = \bigwedge_{i=1}^n \Big| dZ^i_{(I) 0} \Big|^2
F(Z_{(I) 0}, \eta_A)
\end{equation}
where $F(Z_{(I) 0}, \eta_A)$ is the result of the non-zero modes
integration. 
This formula shows that there are two sources of patch dependence: one
in the integration measure and the other in the integral of the
non-zero modes (since the decomposition between zero modes and
non-zero mode depends upon the choice of the patch). 
Therefore, it is convenient to introduce an auxiliary variable $x^i$
and to rewrite the above expression as  
\begin{eqnarray}\label{rapAB}
\omega^{(n|n)}_{(I)} &=& \int \bigwedge_{i=1}^n |dx^i \delta(x^i -
Z^i_{(I)0})|^2 \bigwedge_{i=1}^n \Big| dZ^i_{(I) 0} \Big|^2  F(Z_{(I)
  0}, \eta_A) \nonumber \\ 
& = & \int \bigwedge_{i=1}^n |dx^i |^2  F(x, \eta_A) \bigwedge_{i=1}^n
\Big| dZ^i_{(I) 0} \delta(x^i - Z^i_{(I)0})\Big|^2  
\end{eqnarray}
The Dirac delta function localizes the expression $F(x, \eta_A)$ on a
single patch but, now, the patch dependence is entirely in the last
factor which can be written as  
\begin{align} \label{rapB} 
\omega^{(n|n)}_{(I)} &= \int \bigwedge_{i=1}^n \Big| d x^i \Big|^2
F(x, \eta_A) \int \Big| d B_i \Big|^2  \Big| d \bar C_i \Big|^2 e^{B_i
  (x^i - Z^i_I) + \bar C_i dZ^i + {\rm c.c.}} \nonumber\\ 
& \equiv \int \bigwedge_{i=1}^n  \Big| d x^i \Big|^2   F(x, \eta_A)
\int \Big| d B_i \Big|^2  \Big| d \bar C_i \Big|^2 \, \omega_{(I)}
\nonumber\\ 
& \omega_{(I)} = e^{d[\bar C_i (x^i - Z_I^i)] + {\rm c.c.}} 
\end{align}
where we have introduced $n$ pairs of auxiliary fields $B_i$ and $\bar
C_i$ with the property that $dB_i = 0$, $d \bar C_i = B_i$ and $d
Z^i_I$ is the usual basis of 1-forms ($d$ is automatically nilpotent). 
The function $\bar C_i (x^i - Z^i_I)$ plays the role of a gauge-fixing
fermion with negative ghost number (carried by the $\bar C_i$) and it
selects the patch on which the functional is evaluated. Notice that 
at this point one can substitute the Dirac delta function with a smooth expression. 
Now, our next problem is how to construct a globally defined
expression for $\omega_{(I)}$.

Following \cite{Becchi:1996iq} we study the variation of the
$\omega_I$'s changing the patch and we get  
\begin{align}
\omega_{(I)} - \omega_{(J)} &= e^{d \left( \bar C_i (x^i - Z^i_I)  +
  {\rm c.c.} \right)} - e^{d \left( \bar C_i (x^i - Z^i_J)  + {\rm
    c.c.} \right)} = \nonumber\\ 
& = d \left[ \bar C_i (Z^i_J - Z^i_I) \int^1_0 e^{d \left[ t \bar C_i 
  (x^i - Z^i_I) + (1-t) ( \bar C_i (x^i - Z^i_J) + {\rm c.c.}
  \right]}\right] = d \omega_{(IJ)} \label{rapD} 
\end{align}
this can be easily proven by observing that $d$ acts only on the
factor in front of the integral, since the argument is $d$-closed, and
the actual effect of the differential operator on the prefactor is to
give a total derivative w.r.t. $t$. 
Hence, the difference between $\omega_{(I)}$ and $\omega_{(J)}$ is
$d$-exact and this allows us to employ the descent equations technique
to derive a ladder of forms $\omega_{(I_1 \dots I_p)}$ defined on the
$p$-intersections ${\cal U}_{(I_1 \dots I_p)}$ of the open sets
$\cal{U}_I$. 

The main point is that, in \cite{Becchi:1996yh, Becchi:1996iq}, a
prescription is given on how to compute the complete set of forms
satisfying the descent equations.  
The forms $\omega_{(I_1 \dots I_p)}$ can be straightforwardly written
as the product of a prefactor $A$ times an interpolating action
$S[t_1, \dots, t_p]$ 
\begin{align}
\omega_{(I_1 \dots I_p)} &= A_{I_1 \dots I_p} \times S[t_1, \dots,
t_p] \nonumber \\
A_{I_1 \dots I_p} &= \frac{1}{p!}\sum_{\alpha=1}^p (-1)^{\alpha+1}
\bar C_{i_1} (x^{i_1} - Z^{i_1}_{(I_1)0}) \dots \overbrace{\bar
  C_{i_\alpha} (x^{i_\alpha} - Z^{i_\alpha}_{(I_\a)0})} \dots \bar
C_{i_p} (x^{i_p} - Z^{i_p}_{(I_p)0}) + {\rm c.c.} \\  
S[t_1, \dots, t_p] &= \int^1_0 \left(\prod_{\alpha=1}^p dt_\alpha 
\right) \exp \{d[\sum_{\beta=1}^p t_\beta \bar C_i (x^i -
Z_{I_\beta}^i) + {\rm c.c.}]\} \delta\left(\sum_{\alpha=1}^p t_\alpha
  -1\right)\label{rapE}\nonumber 
\end{align} 
The symbol $\overbrace{\phantom{w}}$ means that the corresponding term
must be omitted in the expression. 
It is easy to verify that they satisfy the descent equations 
\begin{equation}\label{rapF}
d \omega^{n}_{(I)}=0\,, \quad \dots \quad
(\delta \omega^{n-p+1})_{(I_1 \dots I_p)} = d \omega^{n-p}_{(I_1 \dots I_p)}
\,, \quad \dots \quad 
(\delta \omega^{0}_{(I_1 \dots I_n)}) = 0 \,.
\end{equation}

At order $p$ the prefactor contains $p$ powers of the anticommuting
variable $\bar C$ so that, in order for the $\bar C$ integral not to
vanish, $n-p-1$ further powers of $\bar C$, which are extracted from
the exponential, are needed. 
The resulting expression is a $n-p-1$ form with ghost number $p$. 
Notice that assembling the monomials $(x^i - Z^i_I)$ into a matrix
with $n$ rows (the number of independent coordinates on the target
space) and $q$ columns (the number of patches of the atlas) the
prefactor computes the determinant of its minors. 
It may happen that from a given order all the determinants vanish. 
This formula provides a suitable choice for the component of the
ladder in the descent equations (each component is defined up to exact
terms).

Using the collating formula for $\omega_{(I)}$ we can finally write  
the globally defined expression
\begin{equation}\label{rapG}
\omega_{global} = \omega_{(I)} + \sum_{l=1}^{n}(- D'' K)^l
\omega_{(l)} 
\end{equation}
where we have compactly denoted $(- D'' K)^l \omega_{(l)} = (- D''
K)^{I_1} \dots (- D'' K)^{I_l} \omega_{(I I_1 \dots I_l)}$ and each
differential operator $D''$ acts on the entire expression on its
right.

At this point one can integrate over the anticommuting fields $\bar C$
and over the commuting pairs $B_i$ and $x^i$ leading to an expression
which is globally defined providing the correct integrand of
(\ref{transB}). 
Notice that, in this way, one selects the form degree needed in the
integration. By following again \cite{Becchi:1995ik}, inserting the
global form $\omega_{global}$ in the integration, we have 
\begin{eqnarray}\label{rapH}
&&\int \bigwedge_{i=1}^n  \Big| d x^i \Big|^2   F(x, \eta_A) \int
\Big| d B_i \Big|^2  \Big| d \bar C_i \Big|^2 
\,  \omega_{global} \nonumber \\ 
&=& \int \bigwedge_{i=1}^n  \Big| d x^i \Big|^2   F(x, \eta_A) 
\sum_{q=0}^{n}(-1)^q \!\! \sum_{I_0 < I_1< \dots <I_q} 
\int_{{\cal C}_{I_0 \dots I_q}} \omega_{(I_0 \dots I_q)}^{n-q}(x, Z_0)
\end{eqnarray}
where we have explicitly written the dependence of the co-chains
$\omega_{(I_0 \dots I_q)}^{n-q}(x, Z_0)$ upon $x^i$ and $Z_0$, in
order to recall that the remaining integrals are over the coordinates  
(restricted to the intersections) $Z_0$. 
Thus, the present formula takes into account all jumps of the path
integral given by the naive definition. The integration over ${\cal
  C}_{I_0 \dots I_q}$ is the integration over the intersection ${\cal
  U}_{I_0 \dots I_q}$ obtained by covering the complete target
space ${\cal M}$ with the partition of unity. 

The final form of the integral of $\omega_{global}$ depends upon the
choice of the partition of unity. Therefore one can choose it in the
most convenient way (for instance it can be chosen such that the
global form rapidly vanishes at infinity\footnote{For the pure spinor
  integration measure, this property replaces the regularization at
  boundary of the pure spinor space discussed in
  \cite{Berkovits:2006vi} while for the regularization at the poles
  $\lambda \cdot \bar \lambda =0$ the regulator \cite{Grassi:2009fe}
  can be used.} or avoiding possible poles). 
Furthermore, the partition of unity depends upon the complex conjugate
of the coordinates $Z_I^i$ and consequently the global form turns out
to be non-holomorphic. 

\section{Examples}

Here, we discuss some example. For some of them, we provide the
construction of the measure using both the algebraic construction and
the collating formula.

\subsection{Cone in {$\mathbb{C}^3$}}

Let us consider the surface in $\mathbb{C}^3$
\begin{equation}
\label{cone}
Z_0^n + Z_1^n + Z_2^n = 0
\end{equation}
which describes a non-compact complex hypersurface. The algebraic
equation is homogenous and therefore it describes a cone over a compact
hypersurface in $\mathbb{P}^2$.  
According to the previous discussion, the integration measure is given
by  
\begin{equation}
\Omega^{(2|0)} = \frac{1}{2} \frac{\epsilon^{ijk} \bar
  Z_k\,  dZ_i \wedge dZ_j }{\sum_{i=0}^2 \bar Z_i Z_i^{(n-1)}}
\end{equation}
This surface can be covered by 3 patches $\mathcal{U}_i$, such
that $Z_i \neq 0$. An easy computation shows that in $\mathcal{U}_0$,
using the coordinates 
\begin{equation}\label{patA}
\phi^0_{(0)} = \lambda = Z_0\,, ~~~~~~~\quad
\phi^1_{(0)} = \gamma = \frac{Z_1}{Z_0}\,, ~~~~~~\quad
\phi^2_{(0)} = u =\frac{Z_2}{Z_0} = i \sqrt{1 + \gamma^n}
\end{equation}
and choosing a suitable $\bar Z_i$ we get
\begin{equation}
\Omega^{(2|0)} = \frac{\partial \lambda}{\lambda^{(n-2)}}
\frac{\partial \gamma}{u^{(n-1)}} 
\end{equation}
which is holomorphic and nowhere vanishing; moreover, the factor
$\frac{\partial \gamma}{u^{(n-1)}}$ is nonsingular if $n=3$, and
indeed this is the holomorphic form for the CY ${\cal{P}}^3 /
\{\gamma^2 + u^2 + 1 = 0\}$. 

For the same space we construct the measure using the collating
formula as discussed above.  
For that purpose, we organize the coordinates as follows: on the patch
${\cal U}_0$ we use the coordinates given in (\ref{patA}). On the the
patch ${\cal U}_1$ and ${\cal U}_2$ we set
\begin{equation}\label{patB}
\phi^0_{(1)} = \lambda'= Z_1 \,, ~~~~~~~\quad
\phi^1_{(1)} = \gamma' = \frac{Z_0}{Z_1}\,, ~~~~~~\quad
\phi^2_{(1)} = u' =\frac{Z_2}{Z_1} 
\end{equation}
and 
\begin{equation}\label{patC}
\phi^0_{(2)} = \lambda''= Z_2 \,, ~~~~~~~\quad
\phi^1_{(2)} = \gamma'' = \frac{Z_0}{Z_2}\,, ~~~~~~\quad
\phi^2_{(2)} = u'' =\frac{Z_1}{Z_2} 
\end{equation}
the transition functions between the patches ${\cal{U}}_1$ and
${\cal{U}}_0$ are 
\begin{equation}\label{patD}
\lambda'=\lambda \gamma\,, \quad
\gamma' = 1/\gamma\,, \quad
u' = u\gamma\,,
\end{equation}
and those between ${\cal{U}}_2$ and ${\cal{U}}_0$
\begin{equation}\label{patEA}
\lambda''= \lambda u\,, \quad 
\gamma'' = \gamma/u\,. \quad
u'' = 1/u
\end{equation}
Since the coordinate $\phi^2_{(I)}$ is always fixed by the constraint
(\ref{cone}) we can take as independent variables the first two
coordinates $\phi^i_{(I)}$ with $i=0,1$, and therefore we introduce
only for them the corresponding ghost fields $\bar C^i$ and the
Lagrange multipliers $B^i$. 

In the present case we have to compute the following terms: the 
2-forms $\omega^{(2)}_I$, where $I=0,1,2$, the 1-forms
$\omega^{(1)}_{[IJ]}$ on the intersections ${\cal U}_{01}, {\cal
  U}_{12}, {\cal U}_{02}$, and the 0-form $\omega_{012}$ in the triple
intersection ${\cal U}_{012}$. 
We have 
\begin{equation}
\label{copA}
\begin{split}
\omega^{(2)}_I &= e^{B_i (x^i - \phi^i_{(I)}) + \bar C_i d
  \phi^i_{(I)}}\,, \\ 
\omega^{(1)}_{IJ} &= \bar C_i (\phi^i_{(J)} - \phi^i_{(I)}) \int_0^1
\int_0^1 dt_I dt_J e^{d \left[t_I \bar C_i (x^i - \phi^i_{(I)})+t_J
    \bar C_i (x^i - \phi^i_{(J)})\right]} \delta(t_I + t_J -1)\\
\omega^{(0)}_{IJK} &= \bar C_i \bar C_j (\phi^i_{(I)} \phi^j_{(J)} +
\phi^i_{(J)}  \phi^j_{(K)} + \phi^i_{(K)}  \phi^j_{(I)})\\
&\times \int_0^1\int_0^1\int_0^1 dt_I dt_J dt_K e^{d
  \left[\sum_{m=0}^2 t_{I_m} \bar C_i (x^i - \phi^i_{(I_m)})\right]}
\delta\left(\sum_{m=0}^2 t_{I_m} -1\right) \\
\end{split}
\end{equation}
which satisfy the descent equations: 
\begin{equation}\label{copB}
d \omega^{(2)}_I = 0\,, ~~~~~~~~~
(\delta \omega^{(2)})_{IJ} = d \omega^{(1)}_{IJ}\,, ~~~~~~~~
(\delta \omega^{(1)})_{IJK} = d \omega^{(0)}_{IJK}\,, ~~~~~~~~
(\delta \omega^{(0)})_{IJK} = 0\,,
\end{equation}
From these equations, acting with the homotopy operator, we can
reconstruct the global form.  
Using the partition of unity
\begin{align}
\rho_0 = \frac{1}{\Delta}\,, \quad
\rho_1 = \frac{|\gamma|^2}{\Delta} \,, \quad 
\rho_2 = \frac{|u|^2}{\Delta} \,, \quad
\Delta = 1 + |\gamma|^2 + |u|^2
\end{align} 
upon integration over the ghost fields $\bar C^i$, over the auxiliary
fields $B^i$ and over the $t$'s, we get the following global 2--form
\begin{equation}
\label{patE}
\begin{split}
\Omega^{(2|0)} =& \frac{1}{2 \Delta^2} \left[ 2 \Delta \left(1-
    \frac{|\gamma|^2}{\gamma} - \frac{|u|^2}{u^3} \right) + \bar
  \gamma (1+\gamma) \left(\gamma-\frac{1}{\gamma}\right) - \frac{\bar
    u \gamma^2}{u^3} (u^2 - 1) + \right. \\
& - \left. \left( |\gamma|^2 \frac{\bar u \gamma^2}{u^2} - |u|^2 \bar
    \gamma \right)(\gamma+u) \left( \frac{1}{\gamma} -
    \frac{\gamma}{u} \right) \right] \partial \lambda \partial \gamma
\,, 
\end{split}
\end{equation}
The form (\ref{patE}) is globally defined and it is not singular. It
is not holomorphic due to the non-holomorphicity of the partition of
unity. In addition, the ratio between the present formula and the
holomorphic one (in the case of n=3) is a globally defined function on
the hypersurface.

Notice that, in the present example, we have set to zero the action
$S[x]$ and the vertex operators ${\cal O}_A(x, p)$ introduced in the
previous section. This simplified the construction. Without setting
them to zero, the result is definitely more interesting and
complicate.

\subsection{K\"ahler form Collating Formula}

The next example is the construction, via the collating formula, of a
global 2 form for the projective space ${\mathbb P}^{2}$. 
We first will do it in a simplified way by starting from a real form
on the different patches and, in the second place, we will employ the
complete construction given in the previous section by starting from a
holomorphic 2 form on the different patches. The two resulting
expressions will differ for a globally defined function. 

Let us consider $\mathbb{CP}^n$. It can be covered by $n+1$ open sets 
$\mathcal{U}_I$, where homogeneous coordinates $\gamma^{(I)}_J =
\frac{z_J}{z_I}$ can be defined (clearly there are only $n$
independent coordinates, since by definition $\gamma^{(I)}_I = 1$).
In the intersection between two patches $\mathcal{U}_{IJ} \equiv
\mathcal{U}_I \cap \mathcal{U}_J$ the following relation holds:
\begin{equation}
\gamma^{(J)}_K = \frac{\gamma^{(I)}_K}{\gamma^{(I)}_J}
\end{equation}
In any patch $\mathcal{U}_I$ a real 2--form can be defined:
\begin{equation}
\omega_I = \sum_{J=1}^{n+1} |\partial \gamma^{(I)}_J|^2 = \frac{1}{2} 
d \sum_{J=1}^{n+1} \left[ \gamma_J^{(I)} \bar \partial \bar
  \gamma_J^{(I)} - \bar \gamma_J^{(I)} \partial \gamma_J^{(I)} \right]
= \frac{1}{2} d \sum_{J=1}^{n+1} |\gamma_J^{(I)}|^2 \mu_J^{(I)}  
\end{equation}
where
\begin{equation}
\label{defmu}
\mu_J^{(I)} = \frac{\bar \partial \bar \gamma_J^{(I)}}{\bar
  \gamma_J^{(I)}} - \frac{\partial \gamma_J^{(I)}}{\gamma_J^{(I)}}\,. 
\end{equation}
The 1-forms $\mu_J^{(I)}$ have some interesting properties: they are 
$d$--closed and, changing from patch $\mathcal{U}_I$ to patch
$\mathcal{U}_J$, they transform according to
\begin{equation}
\mu_K^{(I)} = \mu_K^{(J)} - \mu_I^{(J)}\,.
\end{equation}
Using the descent equation
\begin{equation}
\left( \delta \omega \right)_{IJ} \equiv \omega_I - \omega_J =
d \omega_{IJ} 
\end{equation}
it is easy to find that
\begin{equation}
\omega_{IJ} = \frac{1}{2} \sum_{K=1}^{n+1} \left[ |\gamma_K^{(I)}|^2
\left( \frac{|\gamma_J^{(I)}|^2 - 1}{|\gamma_J^{(I)}|^2} \right)
\mu_K^{(I)} + \frac{|\gamma_K^{(I)}|^2}{|\gamma_J^{(I)}|^2}
\mu_J^{(I)} \right]
\end{equation}
while, in the triple intersection $\mathcal{U}_{IJK}$, all the
$\omega_{IJK}$ vanish. 
It is useful to take, as partition of unity, the expression
\begin{equation}
\rho^J = \frac{|\gamma^{(I)}_J|^2}{\sum_{K=1}^{n+1}|\gamma^{(I)}_K|^2}
\end{equation}
Plugging everything in (\ref{rapG}), we find that the globally defined  
2--form is 
\begin{equation}
\Lambda = \frac{n+1}{2} d \frac{\sum_J |\gamma_J^{(I)}|^2
  \mu_J^{(I)}}{\sum_K |\gamma_K^{(I)}|^2}
\end{equation}
But since, as it is evident from eq. (\ref{defmu})
\begin{equation}
|\gamma_J^{(I)}|^2 \mu_J^{(I)} = (\bar \partial - \partial)
|\gamma_J^{(I)}|^2 
\end{equation} 
while $d = (\bar \partial + \partial)$, it turns out that 
\begin{equation}\label{KK}
\Lambda = (n+1) \partial \bar \partial \log \sum_{K=1}^{n+1}
|\gamma_K|^2 = (n+1) K
\end{equation}
As we anticipated, we now derive a global 2-form starting from the
2-forms on the three patches  
\begin{align}
\omega_0 = \partial u \partial v, \quad \omega_1 = - \frac{1}{u^3}
  \partial u \partial v, \quad \omega_2 = -\frac{1}{v^3} \partial u
  \partial v 
\end{align}
Using the formula (\ref{rapH}) we can compute the 2-cochain living on the 
simple intersections 
\begin{equation}  
\begin{split}
\omega_{01} &= \frac{1}{2} (1 - \frac{1}{u^2} ) \left[ (1 + u)
  \partial v - v \partial u \right] \\
\omega_{12} &= \frac{1}{2} \left[ ( \frac{1}{u^2} - \frac{1}{uv^2} )
  \partial v - ( \frac{1}{v^2} - \frac{1}{vu^2} ) \partial u \right]
\\ 
\omega_{20} &= \frac{1}{2} (1 - \frac{1}{v^2} ) \left[ (1 + v)
  \partial u - u \partial v \right]
\end{split}
\end{equation}
and finally the 0-form living on the triple intersection  
\begin{align}  
\omega_{012} &= \left( u + v + \frac{1}{uv} - \frac{u}{v} -
  \frac{v}{u} - 1 \right) 
\end{align}
After some simple algebraic manipulation, (\ref{rapG}) can be
rewritten as 
\begin{equation}  
\omega_0^{gl} = \omega_o + d \omega_{20} \rho_2 - d \omega_{01}
  \rho_1 - \omega_{20} d \rho_2 + \omega_{01} d \rho_1 + 2 \omega_{012}
  d \rho_1 d \rho_2 - d \omega_{012} \left( d \rho_1 \rho_2 - d \rho_2
  \rho_1 \right) 
\end{equation}
therefore, by inserting the different pieces we arrive at the cumbersome 
expression  
\begin{align}  
\omega_0^{gl} &= \partial u \partial v + \left( - \frac{|v|^2}{\Delta}
  (1 + \frac{1}{v^3}) - \frac{|u|^2}{\Delta} (1 + \frac{1}{u^3})
  \right) \partial u \partial v + \nonumber \\
& - \frac{1}{2 \Delta^2} \left[ (\frac{1}{u^2} - \frac{1}{uv^2} ) 
  \partial v + (\frac{1}{v^2} - \frac{1}{vu^2}) \partial u \right]
  \left[ |v|^2 (u \bar{\partial} \bar{u} + \bar{u} \partial u) - |u|^2
  (v \bar{\partial} \bar{v} + \bar{v} \partial v) \right] + \nonumber \\
& + \frac{1}{2 \Delta^3} (1 - \frac{1}{u^2}) \left[ (1 + u) \partial v
  - v \partial u \right] \left[ (1 + |u|^2) (v \bar{\partial} \bar{v}
  + \bar{v} \partial v) - |v|^2 (u \bar{\partial} \bar{u} + \bar{u}
  \partial u) \right] + \nonumber\\ 
& - \frac{1}{2 \Delta^3} (1 - \frac{1}{v^2}) \left[ (1 + v) \partial u
  - u \partial v \right] \left[ (1 + |v|^2) (u \bar{\partial} \bar{u}
  + \bar{u} \partial u) - |u|^2 (v \bar{\partial} \bar{v} + \bar{v}
  \partial v) \right] + \nonumber\\ 
& + \frac{2}{\Delta^3} \left(u + v + \frac{1}{uv} - \frac{u}{v} - 
  \frac{v}{u} - 1 \right) \left(u \bar{\partial} \bar{u} + \bar{u}
  \partial u\right) \left(v \bar{\partial} \bar{v} + \bar{v} \partial
  v\right) 
\end{align} 
In order to compare the two results, we have to evaluate $\Lambda
\wedge \Lambda$ (in the case $n=2$), and $\omega_0^{gl} \wedge
\bar{\omega}_0^{gl}$; the results are proportional up to a globally
defined function.

\subsection{Pure Spinors}
\label{sec:ps}

Let us recall some basic fact about the pure spinors. We consider a
spinor $\lambda^\a$ in 10d satisfying the algebraic equation 
$W^m(\lambda) = \lambda^\a \gamma^m_{\a\b} \lambda^\b=0$ 
where $m=0,\dots,9$. This is not a complete intersection since the
constraints $W^m(\lambda)=0$ are not independent from each others. 
The coordinates of the PS space are $\lambda^+$, $u_a$ and $u^{ab}$
($a=1 \dots 5$, and $u^{ab} = - u^{ba}$), constrained by the relation
$u_a = \frac{1}{8} \epsilon_{abcde} u^{bc} u^{de}$ which solve the pure spinor condition. Correspondingly
there are 16 patches: 
$$\underset{1}{\underbrace{\cal{U}_+}}, \quad
\underset{5}{\underbrace{{\cal{U}}_a}}, \quad
\underset{10}{\underbrace{{\cal{U}}_{ab}}}$$
The transition functions allowing us to move from a patch to another
one can be found in \cite{Nekrasov:2005wg}; we only quote those needed
for our purpose, that is those that tell us how to move from the patch
${\cal{U}}_a$ or ${\cal{U}}_{ab}$ to ${\cal{U}}_+$:
\begin{equation}
{\text{in\, \cal{U}}_{+a}}
\begin{cases}
\tilde{\lambda}^+ &= \lambda^+ u_a \\
\tilde{u}^{ij} &= \frac{1}{2} \epsilon_{ijakl} \frac{u^{kl}}{u_a} \\
\tilde{u}^{ai} &= \frac{1}{2} \epsilon_{ijakl} \frac{u^{kl}}{u_a}
\end{cases}, 
\qquad
{\text{in\, \cal{U}}_{+ab}}
\begin{cases}
\tilde{\lambda}^+ &= \lambda^+ u^{ab} \\
\tilde{u}^{ab} &= \frac{1}{u^{ab}} \\
\tilde{u}^{ia} &= \frac{u^{ib}}{u^{ab}}\\
\tilde{u}^{ib} &= \frac{u^{ia}}{u^{ab}} \\
\tilde{u}^{ij} &= u^{ab} + \frac{u^{ib}u^{ja}-u^{jb}u^{ia}}{u^{ab}} 
\end{cases}
\end{equation}
We have to find 11 independent globally defined 1--forms, so that we
can use the collating formula to globalize $d\lambda^+$ and $du^{ab}$.
The result is:
\begin{equation}
\label{eq:lamplgl}
\Lambda_{glob} = d \left( \frac{\lambda^+ \left( 1 + \sum_{a=1}^5
u_a^2 \bar{u}^a + \frac{1}{2} \sum_{a,b=1}^5 {u^{ab}}^2 \bar{u}_{ab}
\right)}{1 + \sum_{a=1}^5 u_a \bar{u}^a + \frac{1}{2} \sum_{a,b=1}^5
u^{ab} \bar{u}_{ab}} \right) 
\end{equation}
and
\begin{equation}
\label{eq:sigmagl}
\Sigma^{ab}_{glob} = d \left( \frac{u^{ab} + u^{ab} u^{ij}
    \bar{u}_{ij} + \frac{1}{2} \left( \bar{u}^a u^{ai} + \bar{u}^b
      u^{bi} \right) u^{jk} \epsilon_{abijk} + \bar{u}_{ab} + u^{ai}
    \bar{u}_{bi} - u^{bi} \bar{u}_{ai} + 2 u^{ai} \bar{u}_{ij}
    u^{jb}}{1 + \sum_{a=1}^5 u_a \bar{u}^a + \frac{1}{2}
    \sum_{a,b=1}^5 u^{ab} \bar{u}_{ab}} \right) 
\end{equation}

The global form is obtained by computing the wedge product of the 11 forms
and by multiplying it with its conjugate
\begin{equation}\label{pofA}
\Omega^{(11)} = 
\Lambda_{glob} \bigwedge_{a<b} \Sigma^{{ab}}_{glob} \wedge \overline{\Lambda_{glob} \bigwedge_{a<b} \Sigma^{{ab}}_{glob}
}
\end{equation}
which should be normalized by $\int \Omega^{(11)}$ to get the correct value for 
physical amplitudes.  
 
 \subsection{Grassmannian $G(2,4)$}
\label{sec:grass}

Grassmannians $G(k,n)$ are the set of $k$--planes in $\mathbb{C}^n$. 
They are an algebraic variety in $\mathbb{CP}^{\left(
    \begin{smallmatrix} n \\ k \end{smallmatrix} \right) - 1}$,  
identified by the Pl\"{u}cker relations. For any $G(k,n)$ we can
define the $\left( \begin{smallmatrix} n \\ k \end{smallmatrix}
\right)$ so--called Pl\"{u}cker coordinates $\lambda_{[i_1 \dots
  i_k]}$ which give explicitly the embedding. For instance, $G(2,4)$
is the algebraic variety in $\mathbb{CP}^5$ given by the relation
$\epsilon^{abcd} \lambda_{ab} \lambda_{cd}$ ($a,b:1 \dots 4$ and
$\lambda_{ba} = - \lambda_{ab}$). It can be covered by 6 patches:
$\mathcal{U}_{ij}$ such that $\lambda_{ij} \neq 0$. \\
In $\mathcal{U}_0$ (where $\lambda_{12} \neq 0$) we can rename the
coordinates as: 
\begin{align}
\lambda_{12} &= \gamma, & \lambda_{13} &= \gamma x, & \lambda_{14} &=
\gamma y \nonumber\\
\lambda_{23} &= \gamma z, & \lambda_{24} &= \gamma w, & \lambda_{34} &=
\gamma t \nonumber
\end{align}
where $t = xw-yz$ is a dependent coordinate, due to the Pl\"{u}cker
relation. After repeating the procedure for all the other patches and
defining the following partition of unity
\begin{equation}
\begin{split}
  \Delta &= 1 + |x|^2 + |y|^2 + |z|^2 + |w|^2 + |t|^2 \\
  \rho_0 &= \frac{1}{\Delta}, \, \rho_x = \frac{|x|^2}{\Delta}, \, 
  \rho_y = \frac{|y|^2}{\Delta}, \\
  \rho_z &= \frac{|z|^2}{\Delta}, \, \rho_w = \frac{|w|^2}{\Delta},
  \, \rho_t = \frac{|t|^2}{\Delta},
\end{split}
\end{equation}
the collating formula can be used to globalize the following 4
independent 1--forms 
\begin{align}
\omega_1 &= d x \rightarrow \Omega_1^{Glob} = d \left[ \frac{x \left(1 
+ \bar{y} + \bar{z} + \bar{t} \right) + \bar{x} + t
\bar{w}}{\Delta}\right] \\
\omega_2 &= d y \rightarrow \Omega_2^{Glob} = d \left[ \frac{y \left(1
+ \bar{w} - \bar{x} + \bar{t} \right) + \bar{y} + t
\bar{z}}{\Delta}\right] \\
\omega_3 &= d z \rightarrow \Omega_3^{Glob} = d \left[ \frac{z \left(1
+ \bar{x} - \bar{w} + \bar{t} \right) + \bar{z} + t
\bar{y}}{\Delta}\right]\\
\omega_4 &= d w \rightarrow \Omega_4^{Glob} = d \left[ \frac{w \left(1
+ \bar{y} + \bar{z} + \bar{t} \right) + \bar{w} + t
\bar{x}}{\Delta}\right]
\end{align}
Alternatively we could have started from the 4--form 
$$\sigma = 4 dx dy dz dw \equiv 4 \Lambda$$
which would have led to the following globally defined 4--form
\begin{equation}
\Sigma = \frac{1}{\Delta^2} \left[ \Delta \left( \frac{\partial
f^x}{\partial x} + \frac{\partial f^y}{\partial y} + \frac{\partial 
f^z}{\partial z} + \frac{\partial f^w}{\partial w} \right) - \left(
\frac{\partial \Delta}{\partial x} f^x + \frac{\partial
  \Delta}{\partial y} f^y + \frac{\partial \Delta}{\partial z} f^z +
\frac{\partial \Delta}{\partial w} f^w \right)\right] \Lambda
\end{equation}
where the $f$ functions are defined to be
\begin{align}
f^x &= x \left( 1 + \frac{|y|^2}{y^4} + \frac{|z|^2}{z^4} +
  \frac{|w|^2}{w^4} \right) - \frac{|x|^2}{3x^3} + \frac{|t|^2}{3 t^3
  w} \\ 
f^y &= y \left( 1 + \frac{|x|^2}{x^4} + \frac{|z|^2}{z^4} +
  \frac{|w|^2}{w^4} \right) - \frac{|y|^2}{3y^3} + \frac{|t|^2}{3 t^3
  z} \\
f^z &= z \left( 1 + \frac{|x|^2}{x^4} + \frac{|y|^2}{y^4} +
  \frac{|w|^2}{w^4} \right) - \frac{|z|^2}{3z^3} + \frac{|t|^2}{3 t^3
  y} \\
f^w &= w \left( 1 + \frac{|x|^2}{x^4} + \frac{|y|^2}{y^4} +
  \frac{|z|^2}{z^4} \right) - \frac{|w|^2}{3w^3} + \frac{|t|^2}{3 t^3
  x} 
\end{align}

\vspace{-.5cm}
 \section{Conclusions}
 
We have used a method developed some years ago in the context of
topological gravity and   topological string theory to construct
globally defined forms. It uses the collating formula and we have
adapted the method to pure spinor string theory and related theories
based on algebraic manifolds. We generalized the Griffiths' residue
method and we derived a solution of the descent equations. Some
examples are presented and the present paper is in preparation of more
physical applications, such as amplitude computations in string theory.  

\vspace{-.5cm}
  \section*{Acknowledgments}
We are grateful to D. Matessi for several discussions on algebraic
geometry.

\end{document}